# Deciphering Cryptic Behavior in Bimetallic Transition Metal Complexes with Machine Learning

*Michael G. Taylor,[1] Aditya Nandy,[1,2] Connie C. Lu[3], and Heather J. Kulik[1,*]*

[1]Department of Chemical Engineering, Massachusetts Institute of Technology, Cambridge, MA 02139

[2]Department of Chemistry, Massachusetts Institute of Technology, Cambridge, MA 02139

[3]Department of Chemistry, University of Minnesota, Minneapolis, MN 55455

AUTHOR INFORMATION

**Corresponding Author**

*email: hjkulik@mit.edu; phone: 617-253-4584

ABSTRACT The rational tailoring of transition metal complexes is necessary to address outstanding challenges in energy utilization and storage. Heterobimetallic transition metal complexes that exhibit metal-metal bonding in stacked "double decker" ligand structures are an emerging, attractive platform for catalysis, but their properties are challenging to predict prior to laborious synthetic efforts. We demonstrate an alternative, data-driven approach to uncovering structure-property relationships for rational bimetallic complex design. We tailor graph-based representations of the metal-local environment for these heterobimetallic complexes for use in training of multiple linear regression and kernel ridge regression (KRR) models. Focusing on oxidation potentials, we obtain a set of 28 experimentally characterized complexes to develop a multiple linear regression model. On this training set, we achieve good accuracy (mean absolute error, MAE, of 0.25 V) and preserve transferability to unseen experimental data with a new ligand structure. We trained a KRR model on a subset of 330 structurally characterized heterobimetallics to predict the degree of metal-metal bonding. This KRR model predicts relative metal-metal bond lengths in the test set to within 5%, and analysis of key features reveals the fundamental atomic contributions (e.g., the valence electron configuration) that most strongly influence the behavior of complexes. Our work provides guidance for rational bimetallic design, suggesting that properties including the formal shortness ratio should be transferable from one period to another.

**TOC GRAPHIC**

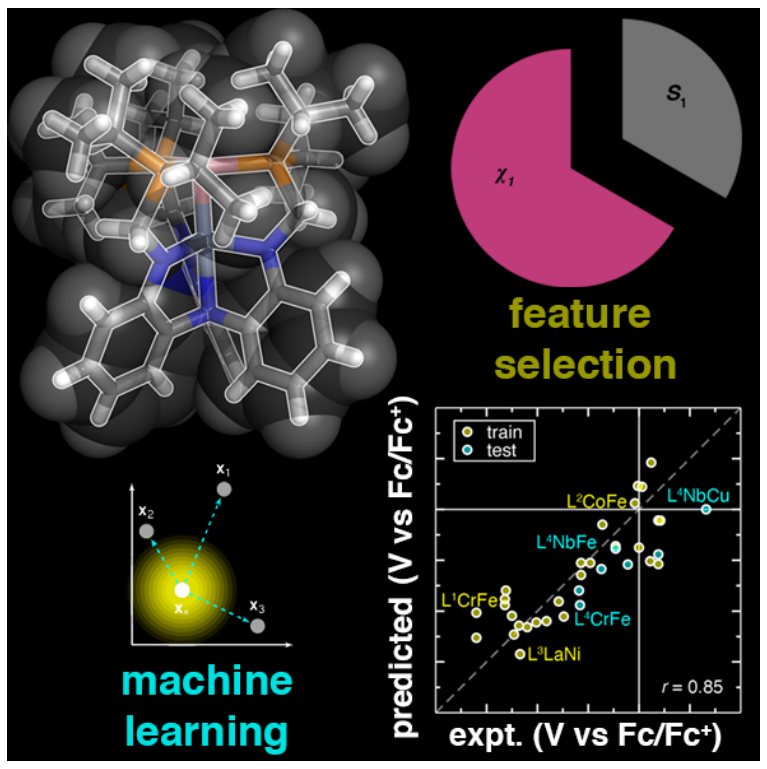

Transition metal complexes inhabit a vast, challenging region of chemical space that holds great promise for the design of functional (e.g., magnetic) materials[1-9] and addressing outstanding challenges in catalysis[10-17] or energy storage[18, 19]. Heterobimetallic complexes are promising because they open a vast chemical space beyond their monometallic counterparts to explore new properties and reactivities. Double decker ligands have been demonstrated[20-22] to constrain two metal atoms (e.g., first-row transition metals[22]) in close proximity, enabling the controlled, combinatorial-like synthesis of a range of metal-metal complexes that exhibit desirable catalytic properties.[23-26] Nevertheless, the rational or first-principles design[27-29] of such complexes is challenged by the complex relationship between the metal identities and their spin state or degree of bonding. In particular, the formal shortness ratio (FSR)[30, 31], which is the ratio of the observed bond length relative to the bond length expected for a single metal-metal bond (i.e., sum of Pauling radii[32]), has been proposed as a measure of strength of bonding observed in complexes that in turn can vary widely depending on differences in metal group number and the *d* electron count of the metals.[22] While the FSR can be used to explain behavior of the complex after characterization, understanding what governs short vs long FSRs would enable better design of complexes. One attractive area for design of heterobimetallic complexes is in tailoring the redox behavior by increasing the number of redox processes and/or tuning their potentials. Toward this end, multiple reversible redox events have been observed to emerge in heterobimetallics when their monometallic counterparts show fewer or none.[22] Tuning the redox potential holds great promise for tailoring the promising catalytic activity of these complexes.[33-35] Nevertheless, guiding principles for design of such redox potentials has remained challenging.

Theoretical (i.e., first-principles) calculations have provided insight into metal-metal bonding[36] and the relationship to redox potentials[37]. Nevertheless, challenges remain in applying low-cost electronic structure methods (i.e., density functional theory or DFT)[38-42] needed to screen large sets (i.e., hundreds) of complexes. In particular, both metal-metal bonding present in these complexes is notoriously difficult to model due to its multiconfigurational nature[42, 43], and numerous challenges for theoretical prediction of redox potentials also remain[44-47]. We were thus motivated by the rich structural and redox data known about these complexes to take a distinct, data-driven approach.

Machine learning (e.g., kernel ridge regression, KRR, or artificial neural network, ANN) models have been trained to predict magnetic properties[48-51], redox and ionization potentials[49, 52, 53] or frontier orbital energies[54], catalyst reaction energetics[55-57], and geometric[48, 49, 58, 59] properties in mononuclear octahedral transition metal complexes as well as metal-organic frameworks[60-62] that contain multiple metals in inorganic secondary building units. Feature selection (e.g., with random forest[48] alone or to preorder features for recursive addition in KRR models[54]) has emerged as a powerful tool for building the structure-property relationships learned on these data sets. A range of graph-based[48, 52, 57, 63-65], geometric[62, 66-68], and even electrostatic[69-72] or quantum mechanical[73-77] descriptors have all been developed for ML model training in inorganic chemistry. Nevertheless, graph based descriptors are particularly attractive for machine learning experimental properties (e.g., stability[78], structure[59], or catalytic activity) because they do not require prior calculation of precise structures that can require high level (i.e., beyond density functional theory accuracy) calculations. Here we turn to tailoring a set of graph-based representations to first predict oxidation and geometric properties of heterobimetallic complexes solely from experimental data. We leverage feature selection to interpret structure-property relationships in heterobimetallics and guide their rational design.

Revised autocorrelations (RACs)[48, 52] are extensions of Moreau-Broto autocorrelation (AC)[64, 65] graph-based descriptors that have been tailored for predictive machine learning in open-shell transition metal chemistry[48]. The standard Moreau-Broto ACs[64, 65] include products (i.e., $P_iP_j$) of heuristic properties of atoms $i$ and $j$ separated by a fixed distance on the molecular graph. RACs are tailored for open-shell transition metal chemistry through the incorporation of property differences:

$$P'_d = \sum_{i}^{start} \sum_{j}^{scope} (P_i - P_j)\delta(d_{ij}, d) \tag{1}$$

Both these and other RACs also include a modified choice of *starting point*, wherein either the metal center(s) or ligand coordinating atom is always included in the sum.[48] Similarly, the *scope* has also been modified, e.g., to be only over specific ligand types[48] or specific components[60] (e.g., in metal organic frameworks, MOFs). Typical heuristics[48] include the number of species an atom is coordinated by (i.e., topology, *T*), simple counting of atoms (i.e., the identity, *I*)[79], nuclear charge (*Z*), covalent radius (*S*), and Pauling electronegativity ($\chi$), but other heuristics such as the effective nuclear charge[80] or atomic polarizability, $\alpha$,[60] have also been employed. When carried out in conjunction with a $d = 3$ depth cutoff, five heuristic properties typically lead to a total feature vector referred to as RAC-155. In machine learning open-shell transition metal chemistry, these RACs have achieved geometry-free prediction of properties including spin splitting energies[48, 49, 81] or redox/ionization potential[48, 53, 54] to within 1-4 kcal/mol of DFT training data. Since RACs do not depend on geometric data, they have been used to predict the metal-ligand bond length, typically to within 0.01-0.03 Å of DFT training data.[48, 49, 58] These models were used to recast experimental ground state spin assignment as a classification problem using an ML model trained[58] to predict the high-spin and low-spin bond lengths of Fe(II) complexes.[59]

We introduce a tailored, sparser representation amenable to machine learning properties (i.e., redox and geometry) of bimetallic transition metal complexes with a limited number of ligand structures (Figure 1). We restrict our start to one of the two metals and compute both products:

$$P_d = \sum_i^{\text{M1 or M2}} \sum_j P_i P_j \delta(d_{ij}, d) \quad (2)$$

as well as differences:

$$P'_d = \sum_i^{\text{M1 or M2}} \sum_j (P_i - P_j) \delta(d_{ij}, d) \quad (3)$$

where the expansion is truncated at $d = 1$ for standard multi-metallic RACs (mm-RACs) or $d = 2$ for extended multi-metallic RACs that also include metal-metal-only products and differences (emm-RACs). While demonstrated here for two metals, the approach could be straightforwardly extended to clusters of other nuclearity. For the heuristic properties, we include the original $Z$, $S$, and $\chi$ along with the number of valence electrons, $n$, and atomic polarizability, $\alpha$ (Supporting Information Table S1). We have excluded topology and the identity because they are invariant over the representative bimetallic complexes (Figure 1). In total, the five properties yield 10 mm-RACs and 30 emm-RACs (Supporting Information Table S2). Analysis of feature-selected RACs, either through minimized cross-validation errors in linear fits[52] or through random-forest-ranked recursive feature addition (RF-RFA, e.g., in kernel ridge regression models)[54], has served as a powerful tool[58] for interpreting structure-property relationships in open-shell transition metal chemistry that we will also demonstrate here for heterobimetallic complexes.

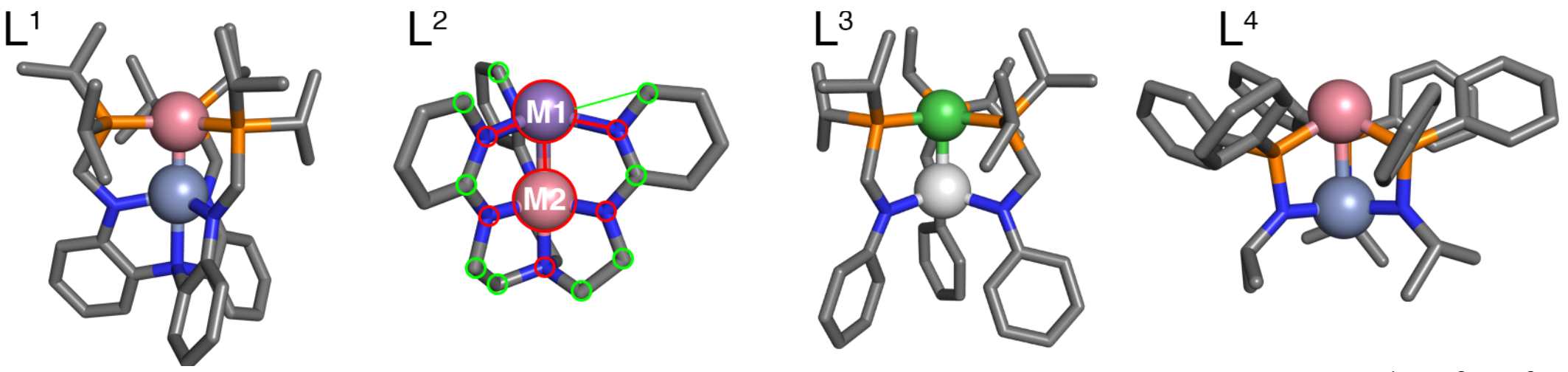

**Figure 1.** Representative ligands studied in this work labeled at top as $L^1$, $L^2$, $L^3$, and $L^4$. The $L^2$ structure has M1 (top metal) and M2 (bottom metal) labeled along with representative components of mm-RACs and emm-RACs (green circles) and emm-RACs only (red circles).

To train a model to predict oxidation potentials, we curated from prior literature reports[22-25, 37, 82-90] a total of 28 bimetallic complexes with three ligand types (i.e., $L^1$,[22, 24, 25, 82-88] $L^2$,[89, 90] and $L^3$ complexes[23, 37], Figure 1 and Supporting Information Table S3). These ligands consist of equatorial N,N- (i.e., $L^2$) or P,N-coordinating (i.e., $L^1$ or $L^3$) atoms in the presence (i.e., $L^1$ and $L^2$) or absence (i.e., $L^3$) of a distal axial coordinating atom. These "double decker" $L^x$M2M1 complexes are typically characterized by complex redox behavior that is challenging to rationalize even within an isostructural series.[23] The upper metal, M1, which is supported by "soft" donors, is believed to undergo the first oxidation process, where the potential is strongly influenced by the lower metal, M2, that has "hard" amide donors.[23] As a result of the "soft" donors around M1, M1 is typically favored to be a later metal in the heterobimetallics, making it the expected site for the first oxidation.

Over this set of 28 complexes, we train a multiple linear regression (MLR) model on a set of mm-RACs selected by the mean absolute error obtained from 10-fold cross validation (CV, Figure 2). The final model has a good correlation with experimental results ($r$ = 0.85) and reasonable fitting error of 0.26 V (Figure 2 and Supporting Information Table S4). Of the ten possible features, only two are selected to minimize CV error on this set, in both cases the 1-depth product of the heuristic of the metals with the surrounding nearest neighbor atoms. The specific heuristics selected are the covalent radius (i.e., $S_1$) or the electronegativity ($\chi_1$) in roughly 1/3 and 2/3 weight, respectively after scaling of the two properties (Supporting Information Figure S1). The largest errors of prediction for the model are challenging cases (e.g., $L^2$FeMn and $L^1$FeFe) where the interplay of metal-metal bonding and spin coupling could be expected to play a role (Supporting Information Table S3). In both cases, the model predicts the oxidation potential to be lower than the observed value by around 0.6-0.7 V (Figure 2 and Supporting Information Table S3). For most other cases, training errors are significantly lower (typically less than 0.2-0.3 V), and model prediction accuracy is good (Figure 2).

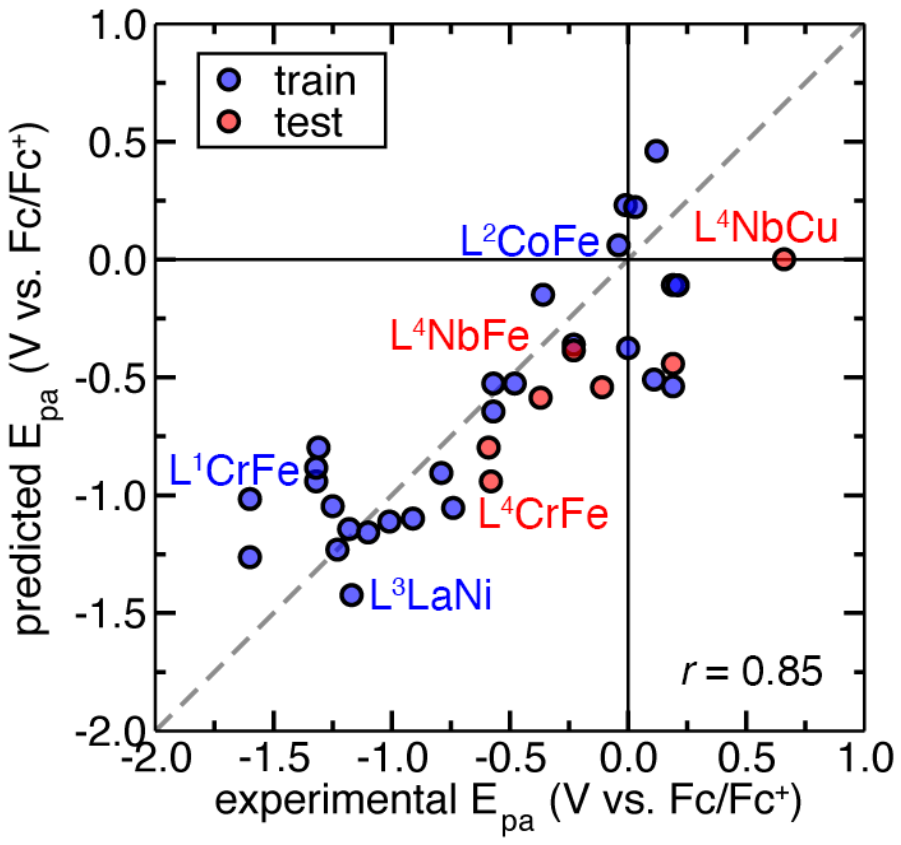

**Figure 2.** Experimental vs predicted oxidation potential ($E_{pa}$, V relative to $Fc/Fc^+$) for training set complexes (blue circles) and test complexes (red circles). The training set correlation is indicated in inset. A gray dashed parity line is shown as, and black zero axes are also shown. Select train and test complexes are indicated in inset by their ligand, bottom metal, M2), and top metal, M1 (i.e., $L^x$M2M1).

To confirm the generality of our approach, we test on set-aside literature results[91, 92] for seven additional complexes with a distinct ligand structure, $L^4$, with N,P-coordination that was not in the training data and include previously unseen (i.e., Nb) elements in the M2 site (Figure 1 and Supporting Information Table S5). Across this set of seven new complexes, the MAE of our model is only slightly worse at around 0.38 V (Figure 2 and Supporting Information Table S3). Generally, errors are lower for the cases where the metal pair had been seen before (i.e., $L^4$CrM1, M1 = Fe, Co, or Cu), highlighting the dominant role of the metal and direct coordination environment in capturing the oxidation potentials of these complexes.

For cases that are challenged by linear models, we can anticipate that more flexible (e.g., KRR) models paired with a larger feature vector and data set could improve in predictive accuracy. Furthermore, properties of these heterobimetallic complexes are believed to be closely related to the relative bond length (i.e., FSR) of the metal-metal bond[22], a quantity that can be readily extracted from experimental structures (i.e., in the Cambridge Structural Database or CSD[93]). Thus, we curated a larger set of all structurally characterized heterobimetallic compounds from the CSD[93] (see *Computational Details*). From over 150,000 candidate complexes, we identified 330 structures that could contain some degree of metal-metal bonding between two distinct metals and were in structures resembling those in our redox set (Figure 3 and Supporting Information Table S6 and see *Computational Details*). The FSR values span a wide range from as low as 0.78 to as high as around 1.30, where an FSR value of 1 corresponds to the expected bond length of a covalent bond, and values less than 1 are expected[21, 22, 30, 31] to exhibit significant, multiple metal-metal bonding (Figure 3 and Supporting Information Table S7). A large number of the complexes have axial and equatorial P and N coordination or otherwise lack a distal axial ligand, exhibiting similarity in the direct coordination environment to our smaller oxidation potential set (Figure 2). Nevertheless, others have coordinating atoms such as S, O, or Cl that were missing in our prior set (Figure 3).

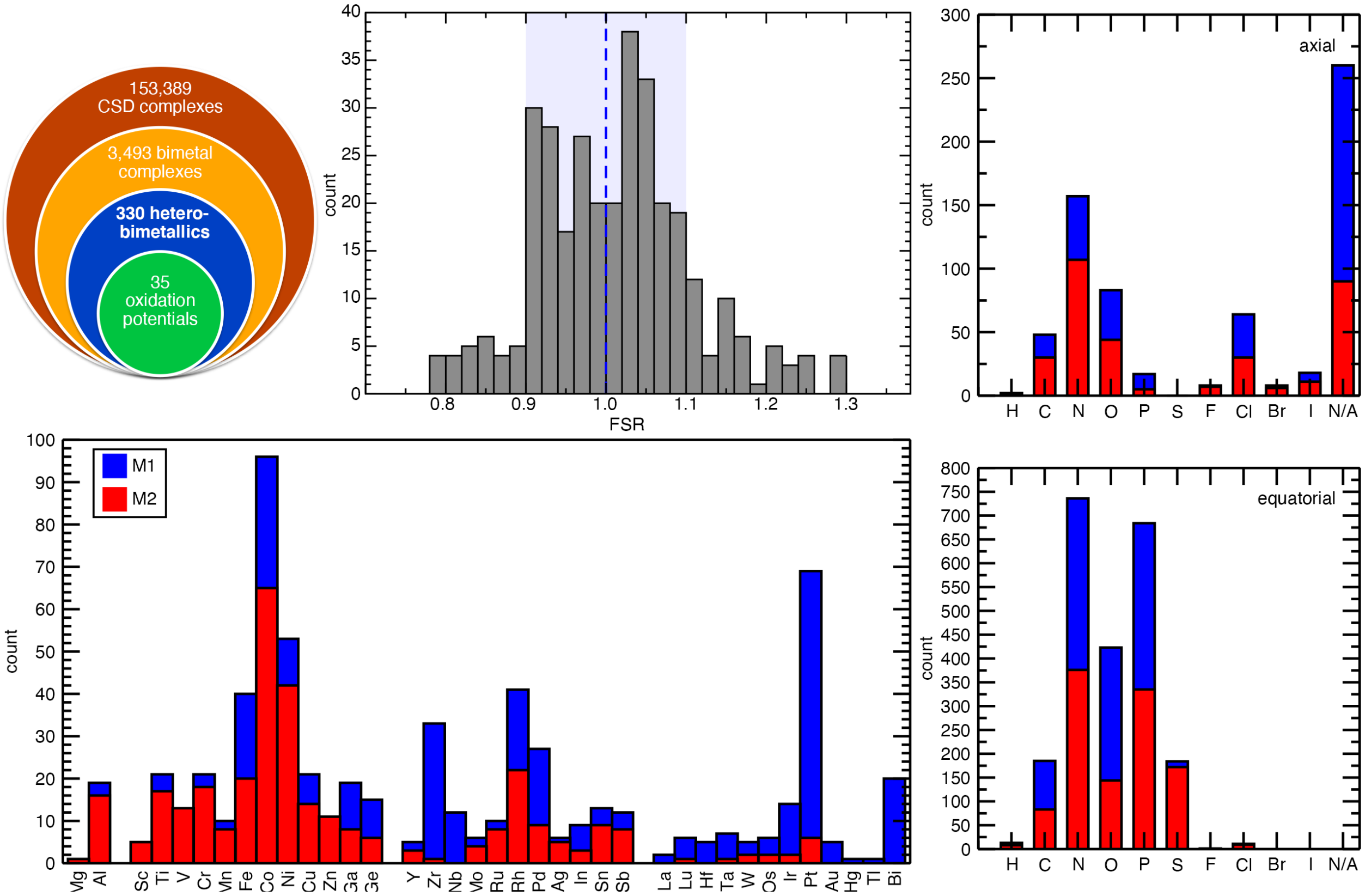

**Figure 3.** Filtering produces set of 330 bimetallic complexes (top left) from the CSD. For these 330 complexes, statistics of the formal shortness ratio (FSR, top middle), element counts for M1 (blue) and M2 (red) are shown (bottom left) along with the total count of elements for coordinating atoms that are not metals in axial positions (top right) or equatorial positions (bottom right).

To train a model to predict the FSR with good performance and to identify important features for predicting FSR, we used random forest-ranked recursive feature addition (RF-RFA) to train a KRR model to predict the FSR across the 330-structure set. We employed an 80/20 train/test split in our model training and feature selection is carried out from subsets of the emm-RACs set (see *Computational Details* and Supporting Information Figure S2 and Table S8). The final model is trained on nine retained emm-RACs and achieves reasonably balanced train (MAE = 0.030) and test (MAE=0.057) errors (Figure 4 and Supporting Information Table S9). The largest errors are comparable in both train and test sets and generally correspond to an underestimation by the model of the longest experimentally measured FSRs that indicate weak bonding (e.g., FSR > 1.1, Figure 4).

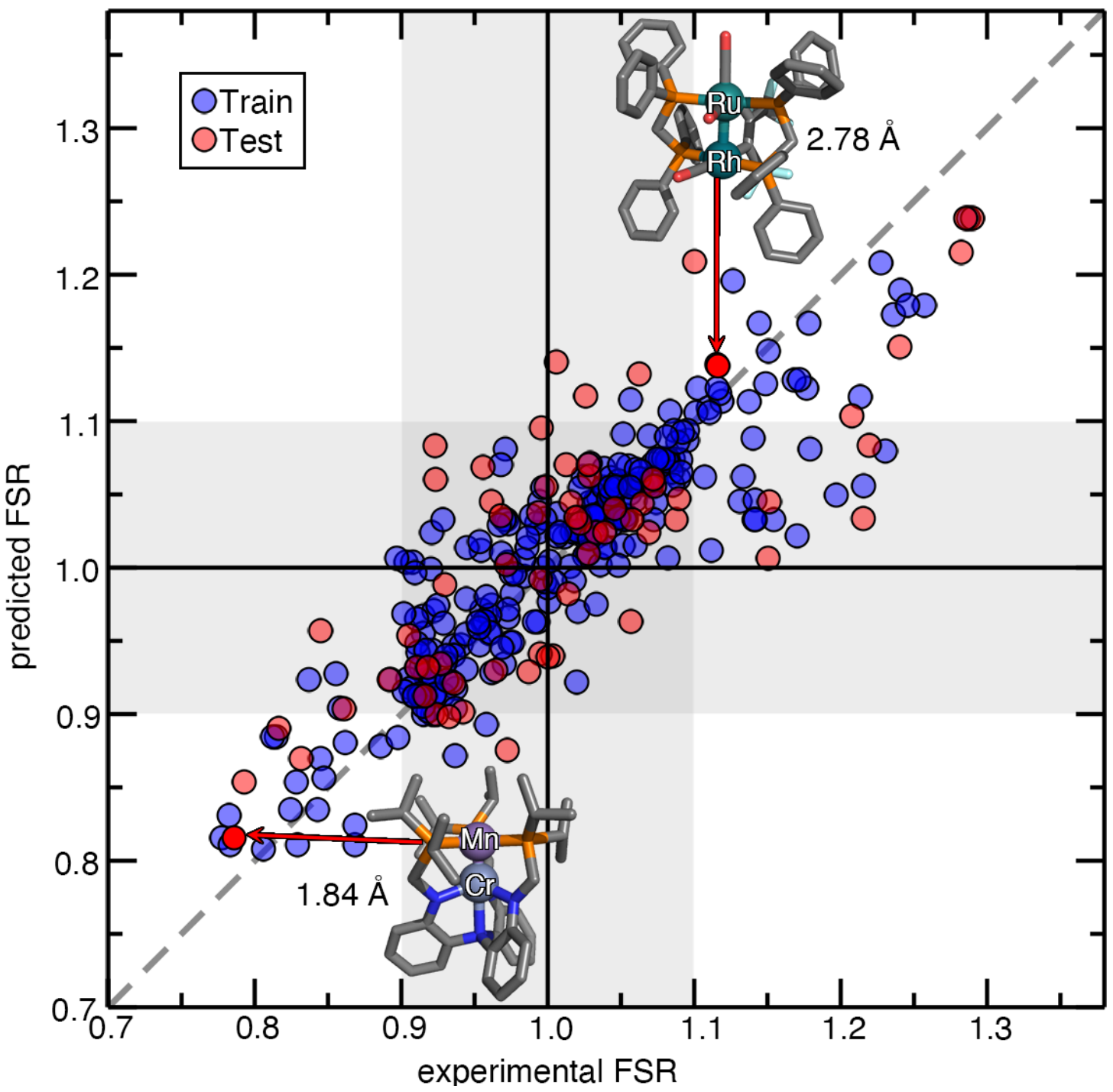

**Figure 4.** Performance of the RF-RFA/KRR model trained on emm-RACs to predict the formal shortness ratio. An 80/20 train/test split was used with training points shown as blue translucent circles, and test points are shown as red translucent circles. A gray dashed parity line is shown, and black bars indicate an FSR of 1. Light gray shaded regions represent the FSR value range from 0.9 to 1.1. Representative test set complexes are labeled in inset with the lowest formal shortness ratio in a CrMn complex (refcode: JULNIP[83]) and a high formal shortness ratio in a RhRu complex (refcode: KADWAO[94]), as indicated in inset.

To understand if there is a type of most challenging cases for structure prediction for our model, we also considered the prediction of the FSR as a classification problem. In line with typically observed FSR values[21, 22, 30, 31], we define an FSR < 0.9 as a short, relatively strong interaction, FSR between 0.9 and 1.1 as intermediate, and then designate FSR > 1.1 as a weak interaction. Based on these designations, the KRR model correctly classifies 239 training complexes (i.e., over 90%) and 53 of the test complexes (i.e., 80%, Figure 4 and Supporting Information Table S10). Of the 13 test structures with prediction errors that would lead to a misclassification of the weak or strong metal-metal interaction in test data, they are equally distributed in the cases where one (i.e., actual or predicted) resides on the boundary whereas the other (i.e., predicted or actual) is classified one of the other (i.e., stronger or weaker) categories (Supporting Information Table S10). Further improvement to the model could likely be achieved by augmenting the data set with additional structures, e.g., through first-principles DFT calculations in multiple spin state configurations and with representations/models that are amenable to predicting[59] spin-state dependent FSR values.

Given the predictive capabilities of the two models for predicting oxidation state and FSR values, we compare what (e)mm-RAC features are most essential for predicting the two quantities in our ML models. For the linear model with mm-RAC coefficients over the smaller oxidation state set, the covalent radius of the metal and its nearest neighbors were key along with the electronegativity of the metal's nearest neighbors (Figure 5). This leads the model to predict oxidation potentials for set-aside test $L^4$M2M1 complexes[91, 92] to be of the order of Fe ~ Co < Cu for M1 in either M2 = Cr or Nb cases because across this series both covalent radius decreases and electronegativity increases, in qualitative accordance with chemical trends (Supporting Information Table S5). The model also correctly predicts that the difference between Fe and Cu M1 complex oxidation potentials will be larger in the case where M2 = Nb than Cr in part because Nb is significantly larger (i.e., in the $S_1$ contribution to the linear model) despite having a comparable electronegativity to Cr (Figure 1 and Supporting Information Table S1). Given the coefficients in the model, the lowest (i.e., most negative) oxidation potentials are observed for high covalent radius elements (e.g., $L^3$ScNi[23]), but electronegativity can play a dominant role, explaining the lower (i.e., more negative) oxidation potential in $L^1$VFe[84] despite the smaller covalent radius (Figure 5).

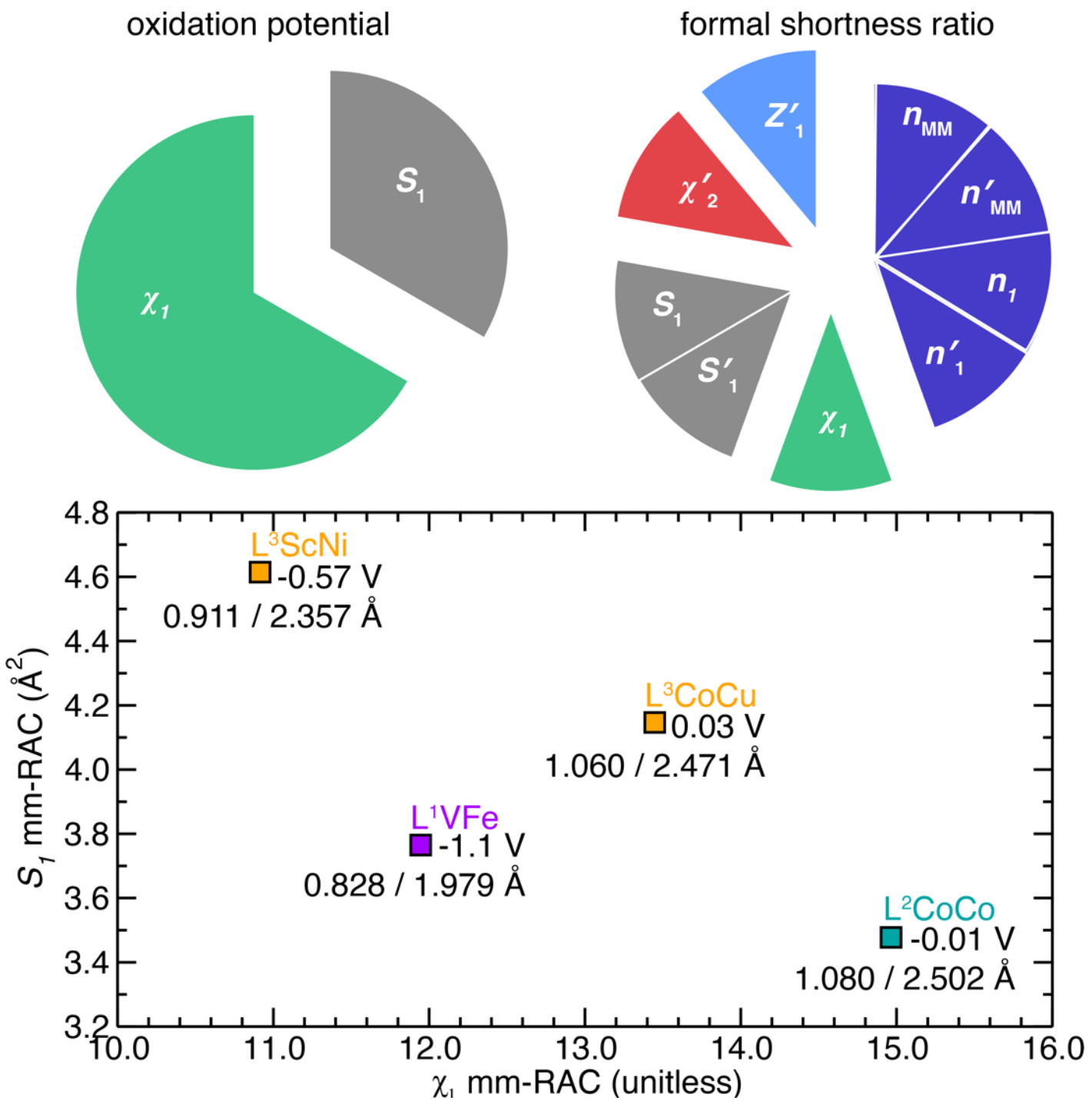

**Figure 5.** (top) Summary of features selected to predict formal shortness ratio and oxidation potential in bimetallic complexes. The oxidation potential linear features are weighted by their representation in the linear model, whereas the FSR features are all shown with equal weight. (bottom) Four representative complexes labeled according to inset from the 28 training complexes in the oxidation potential linear model are shown with their respective mm-RAC descriptors on each axis. The resulting relative oxidation potential and formal shortness ratio and metal-metal distances are labeled on each point, which are colored by ligand scaffold.

Comparing this feature set to those selected to predict FSR in our RF-RFA/KRR model trained on emm-RACs, we note that these two features for predicting oxidation potential are indeed retained (Figure 5). Beyond those two features, a strong dependence on the number of valence electrons also emerges, either in the form of the product of the two metals' valence electron count or in the nearest neighbors to the metals, as indicated by its inclusion in four of the selected features (Figure 5). The only selected feature that depends on the second nearest neighbors to the metal is the atomic polarizability, possibly indirectly encoding through-bond effects otherwise missed in our model (Figure 5). While more refined treatment of more distant features could be useful in improving regression accuracy, this observation is consistent with analysis of important features for predicting bond lengths in mononuclear systems that were observed to be balanced both between metal proximal and more distant features as well as both geometrically and electronically derived heuristic properties.[48]

Finally, only a single nuclear-charge-dependent feature, the difference among the metals and first nearest neighbors, appears in our nine-feature set (Figure 5). Taken together with the relative emphasis on the number of valence electron descriptors, we conclude that electron configuration and group are more important than period effects on the FSR. While it is intuitive that normalized bond lengths might depend less on the principal quantum number than the absolute valence states would, this result nevertheless suggests that principles achieved on one set of metal-metal bonds in readily synthesized complexes (e.g., between first row metals) should be transferable to other rows of the periodic table. For example, a training complex with 4*d*/5*d* M2 = Mo and M1 = W coordinated by N atoms (FSR = 0.82, refcode: GUGDOC[95]) has a comparable FSR to the shortest, nearly-isovalent and accurately predicted test set 3*d* M2 = Mn and M1 = Cr complex coordinated by N and P atoms (FSR = 0.79, refcode: JULNIP[83], see Figure 4). Similar trends are apparent in isovalent heterobimetallic complexes of later transition metals: the 5*d* training complex with M2 = Os M1 = Ir (FSR: 1.11, refcode: AQEYIF[96]) has comparable ligand chemistry (i.e., C and P coordinating atoms) as well as FSR to the accurately predicted 4*d* M2 = Ru, M1 = Rh isovalent test set complex (FSR: 1.12, refcode: KADWAO[94], see Figure 4). Nevertheless, synthetic challenges would remain in using these observations for design of oxidation potential or FSR, as heavier metals are more frequently installed in the upper, M1 site, whereas lighter metals have occurred more frequently in the lower, M2 site, preventing scaffold-independent design of metal-metal pairs (Figure 3). Notably, when the period of the metals are held fixed, FSR and oxidation potential vary consistently, with low FSR corresponding to more negative oxidation potential, indicating limited opportunities for independent tuning of the two properties for independent control of properties (Figure 5).

In summary, we have demonstrated a data-driven approach to uncovering and predicting trends in key properties, the first oxidation potential and the formal shortness ratio, that are governed by ligand-directed metal-metal interactions in bimetallic complexes. We have shown how a series of sparse, graph-based descriptors enable the development of transferable linear models to predict the oxidation potential on unseen ligand structures and metal combinations. We then extracted all compatible bimetallic complexes that had been structurally characterized in the Cambridge Structural Database, for a total of 330 diverse transition metal complexes. Over this set, we showed how a slightly larger feature vector enabled prediction of the formal shortness ratio as a key indicator of degree of metal-metal interaction. Good train and test set performance as well as a low rate (10-20%) of misclassification, which occurred only in cases where either the predicted or actual FSR was borderline, suggested the model had learned the key underlying trends that predict metal-metal interactions.

Analysis of the features selected for these models confirmed the primary effect of relative size of the metal and nearest neighbor atoms for both properties along with the relatively high importance of valence electron configuration over the larger set. While our effort represents the first attempt to tackle such a formidable problem with machine learning, we expect the observations to be general to other applications of bimetallic complexes, including their design as catalysts. Importantly, the approach used here was general not only to transition metals but to main group and lanthanide metals that could be beneficial in far ranging applications beyond catalysis, and this study presents the theoretical foundation for a larger mapping of the as-yet unknown possible materials for further experimental study.

*Computational Details.* A set of bimetallic complexes was curated from the Cambridge Structural Database (CSD) using the Conquest graphical interface and the Python API, in all cases applied to the CSD v5.41 from the August 2020 data set.[93] A query for molecules containing exactly two metals was carried out producing an initial set of 153,589 complexes that were then refined to 3,493 unique bimetallic structures that contained only two metals with coordination numbers between 4 and 6. To ensure some rigidity in the ligand structure and correspondence to the oxidation potential set, no more than five ligands were permitted per complex, sandwich complexes were excluded, complexes consisting only of monodentate ligands were excluded, and at least two bonds between components were required besides the metal-metal bond (Supporting Information Tables S6 and S11). Further, complexes were removed with higher coordination numbers (>8), unexpected denticities, or very long FSR values (>1.5), resulting in 330 unique structures for the FSR data set (Supporting Information Tables S6 and S11). All 35 oxidation potential set molecules are included in this set, which was further refined by searching for abstracts associated with structures that contained cyclic voltammetry keywords using the pybliometrics code[97]. This final oxidation potential set consisted of 35 complexes[22-25, 37, 82-92], 7 of which were held out as set-aside test data[91, 92].

Both multiple-linear regression (MLR) and kernel ridge regression (KRR) models were trained in scikit-learn[98]. The (e)mm-RAC features were applied to the CSD structures using the molSimplify[54, 99] code[100]. Ten-fold cross validation on the mean absolute error was used to select features for the MLR model from mm-RACs on the 28-complex oxidation potential set (Supporting Information Figure S1). A KRR model with a Gaussian kernel was trained to predict the formal shortness ratio from emm-RAC features normalized to have zero mean and unit variance (Supporting Information Table S7). Random-forest-ranked recursive feature addition (RF-RFA)[54] was applied to the full 330 dataset, with features added only if they reduced the 10-fold cross validation MAE (Supporting Information Figure S2). The training data was further split to 80/20 train and validation, on which the random search of hyperparameters was applied based on 10-fold cross-validation MAE to produce the final KRR model (Supporting Information Table S8).

The complete data set and reasons for excluding points from the CSD along with features of retained data and the model hyperparameters are provided in the Supporting Information.

ASSOCIATED CONTENT

**Supporting Information**. The following files are available free of charge. Heuristic atomic properties in (e)mm-RACs; Enumeration of properties in (e)mm-RACs; Relative oxidation potentials and properties for training data set; Cross-validation error for model selection on oxidation potential; Train and test performance statistics for multiple linear regression model; Relative oxidation potentials and properties for test set molecules; Details of CSD structure curation steps; Additional radii used for formal shortness ratio evaluation; KRR model feature selection; Hyperparameters for KRR model; Train and test performance statistics for RF-RFA/KRR model; Statistics on classification of FSR. (PDF)

Details and refcodes of 35-complex oxidation potential set; details and refcodes of 330-complex CSD set; KRR model and feature vector data. (ZIP)

AUTHOR INFORMATION

**Notes**

The authors declare no competing financial interest.

ACKNOWLEDGMENT

This work is supported as part of the Inorganometallic Catalysis Design Center, an Energy Frontier Research Center funded by the U.S. Department of Energy, Office of Science, Basic Energy Sciences under Award DE-SC0012702. A.N. was partially supported by a National Science Foundation Graduate Research Fellowship under Grant #1122374. Database construction and infrastructure was partially supported by the Office of Naval Research under grant number N00014-20-1-2150. H.J.K. holds a Career Award at the Scientific Interface from the Burroughs Wellcome Fund and an AAAS Marion Milligan Mason Award, which supported this work. The authors thank Adam H. Steeves for providing a critical reading of the manuscript.

**Supporting Information for**

***Deciphering Cryptic Behavior in Bimetallic Transition Metal Complexes with Machine Learning***

Michael G. Taylor,[1] Aditya Nandy,[1,2] Connie C. Lu[3], and Heather J. Kulik[1,*]
[1]Department of Chemical Engineering, Massachusetts Institute of Technology, Cambridge, MA 02139
[2]Department of Chemistry, Massachusetts Institute of Technology, Cambridge, MA 02139
[3]Department of Chemistry, University of Minnesota, Minneapolis, MN 55455

**Contents**

**Table S1.** Heuristic atomic properties in (e)mm-RACs by element for all metals and relevant coordinating atoms studied in this work: nuclear charge ($Z$), number of valence electrons ($n_{ve}$), covalent radius (S), atomic polarizability ($\alpha$), and Pauling electronegativity ($\chi$).

| | $Z$ | $n_{ve}$ | $S$ | $\alpha$ | $\chi$ |
|---|---|---|---|---|---|
| H | 1 | 1 | 0.37 | 4.50711 | 2.2 |
| Li | 3 | 1 | 1.33 | 164.1125 | 0.98 |
| B | 5 | 3 | 0.85 | 20.5 | 2.04 |
| C | 6 | 4 | 0.77 | 11.3 | 2.55 |
| N | 7 | 5 | 0.75 | 7.4 | 3.04 |
| O | 8 | 6 | 0.73 | 5.3 | 3.44 |
| F | 9 | 7 | 0.71 | 3.74 | 3.98 |
| Na | 11 | 1 | 1.55 | 162.7 | 0.93 |
| Mg | 12 | 2 | 1.39 | 71.2 | 1.31 |
| Al | 13 | 3 | 1.26 | 57.8 | 1.61 |
| Si | 14 | 4 | 1.16 | 37.3 | 1.9 |
| P | 15 | 5 | 1.06 | 25 | 2.19 |
| S | 16 | 6 | 1.02 | 19.4 | 2.58 |
| Cl | 17 | 7 | 0.99 | 14.6 | 3.16 |
| Sc | 21 | 3 | 1.7 | 97 | 1.36 |
| Ti | 22 | 4 | 1.36 | 100 | 1.54 |
| V | 23 | 5 | 1.22 | 87 | 1.63 |
| Cr | 24 | 6 | 1.27 | 83 | 1.66 |
| Mn | 25 | 7 | 1.39 | 68 | 1.55 |
| Fe | 26 | 8 | 1.25 | 62 | 1.83 |
| Co | 27 | 9 | 1.26 | 55 | 1.88 |
| Ni | 28 | 10 | 1.21 | 49 | 1.91 |
| Cu | 29 | 11 | 1.38 | 46.5 | 1.9 |
| Zn | 30 | 12 | 1.31 | 38.67 | 1.65 |
| Ga | 31 | 3 | 1.24 | 50 | 1.81 |
| Ge | 32 | 4 | 1.21 | 40 | 2.01 |
| Br | 35 | 7 | 1.14 | 21 | 2.96 |
| Y | 39 | 3 | 1.63 | 162 | 1.22 |
| Zr | 40 | 4 | 1.54 | 112 | 1.33 |
| Nb | 41 | 5 | 1.47 | 98 | 1.6 |
| Mo | 42 | 6 | 1.38 | 87 | 2.16 |
| Ru | 44 | 8 | 1.25 | 72 | 2.2 |
| Rh | 45 | 9 | 1.25 | 66 | 2.28 |
| Pd | 46 | 10 | 1.2 | 26.14 | 2.2 |
| Ag | 47 | 11 | 1.28 | 55 | 1.93 |
| In | 49 | 3 | 1.42 | 65 | 1.78 |
| Sn | 50 | 4 | 1.4 | 53 | 1.96 |
| Sb | 51 | 5 | 1.4 | 43 | 2.05 |
| I | 53 | 7 | 1.4 | 32.9 | 2.66 |
| La | 57 | 3 | 1.69 | 215 | 1.1 |
| Lu | 71 | 3 | 1.62 | 137 | 1.27 |
| Hf | 72 | 8 | 1.5 | 103 | 1.3 |
| Ta | 73 | 5 | 1.38 | 74 | 1.5 |
| W | 74 | 6 | 1.46 | 68 | 2.36 |
| Os | 76 | 8 | 1.28 | 57 | 2.2 |
| Ir | 77 | 9 | 1.37 | 54 | 2.2 |
| Pt | 78 | 10 | 1.23 | 48 | 2.28 |
| Au | 79 | 11 | 1.24 | 36 | 2.54 |
| Hg | 80 | 2 | 1.49 | 33.91 | 2 |
| Tl | 81 | 3 | 1.44 | 50 | 1.62 |
| Bi | 83 | 5 | 1.51 | 48 | 2.02 |

**Table S2.** Enumeration of properties in (e)mm-RAC featurizations in terms of property whether it is metal-centered or metal-metal only and property included. The notation used in the main text indicates the depth as a subscript (i.e., 1, 2, or mm) and ' is used for differences. The 1-depth mc RACs are the ones used in predicting oxidation potential.

| | metal-metal-only RAC | 1-depth mc RAC | 2-depth mc RAC |
|---|---|---|---|
| *Z* | 1 product + 1 difference | 1 product + 1 difference | 1 product + 1 difference |
| *S* | 1 product + 1 difference | 1 product + 1 difference | 1 product + 1 difference |
| $\chi$ | 1 product + 1 difference | 1 product + 1 difference | 1 product + 1 difference |
| *n* | 1 product + 1 difference | 1 product + 1 difference | 1 product + 1 difference |
| $\alpha$ | 1 product + 1 difference | 1 product + 1 difference | 1 product + 1 difference |
| Total | 10 | 10 | 10 |

**Table S3.** Relative oxidation potentials (V vs. Fc/$Fc^+$ in V) for 28 complexes from literature references[1-14] for complexes of the form $L^xM2M1$ where $L^X$ is the ligand type described in the main text, M1 is the upper metal, and M2 is the lower metal. The predicted value of the multiple linear regression model (i.e., training error) is also shown in V with the signed error (i.e., predicted-actual) and absolute value of the error, both in V. The largest prediction error is bolded. The fitting expression is $E_{pa} = 0.409*S_1 + 0.782*\chi_1 - 0.641$ in zero mean and unit variance scaled mm-RACs or $E_{pa} = 1.3143*S_1 + 0.58528*\chi_1 - 13.098$ without scaling. The descriptor values are provided in the Supporting Information .zip file, and the atomic heuristic properties entering into the 1-depth RACs are provided in Supporting Information Table S2.

| Ligand | M2 | M1 | actual | predicted | signed error | abs error |
|---|---|---|---|---|---|---|
| $L^1$ | Al | Ni | -0.74 | -1.05 | -0.31 | 0.31 |
| $L^1$ | Cr | Co | -1.31 | -0.80 | 0.51 | 0.51 |
| $L^1$ | Cr | Cr | -1.60 | -1.26 | 0.34 | 0.34 |
| $L^1$ | Cr | Fe | -1.32 | -0.94 | 0.38 | 0.38 |
| $L^1$ | Cr | Mn | -1.18 | -1.14 | 0.04 | 0.04 |
| $L^1$ | Cr | Ni | -1.32 | -0.88 | 0.44 | 0.44 |
| $L^1$ | Fe | Co | -0.23 | -0.36 | -0.13 | 0.13 |
| $L^1$ | Fe | Fe | 0.11 | -0.51 | -0.62 | 0.62 |
| $L^1$ | Ti | Co | -0.79 | -0.91 | -0.12 | 0.12 |
| $L^1$ | Ti | Fe | -1.25 | -1.05 | 0.20 | 0.20 |
| $L^1$ | V | Co | -1.60 | -1.02 | 0.58 | 0.58 |
| $L^1$ | V | Fe | -1.10 | -1.16 | -0.06 | 0.06 |
| $L^1$ | V | Ni | -0.91 | -1.10 | -0.19 | 0.19 |
| $L^2$ | Co | Co | -0.01 | 0.23 | 0.24 | 0.24 |
| $L^2$ | Co | Fe | -0.04 | 0.06 | 0.10 | 0.10 |
| $L^2$ | Co | Mn | 0.00 | -0.38 | -0.38 | 0.38 |
| $L^2$ | Fe | Fe | 0.19 | -0.11 | -0.30 | 0.30 |
| **$L^2$** | **Fe** | **Mn** | **0.19** | **-0.54** | **-0.73** | **0.73** |
| $L^1$ | Ga | Ni | -0.57 | -0.53 | 0.04 | 0.04 |
| $L^1$ | In | Ni | -0.36 | -0.15 | 0.21 | 0.21 |
| $L^1$ | Al | Rh | 0.21 | -0.11 | -0.32 | 0.32 |
| $L^1$ | Ga | Rh | 0.12 | 0.46 | 0.34 | 0.34 |
| $L^3$ | Co | Cu | 0.03 | 0.22 | 0.19 | 0.19 |
| $L^3$ | Sc | Ni | -0.57 | -0.64 | -0.07 | 0.07 |
| $L^3$ | Y | Ni | -1.23 | -1.23 | 0.00 | 0.00 |
| $L^3$ | La | Ni | -1.17 | -1.42 | -0.25 | 0.25 |
| $L^3$ | Lu | Ni | -1.01 | -1.11 | -0.10 | 0.10 |
| $L^3$ | Ga | Ni | -0.48 | -0.53 | -0.05 | 0.05 |

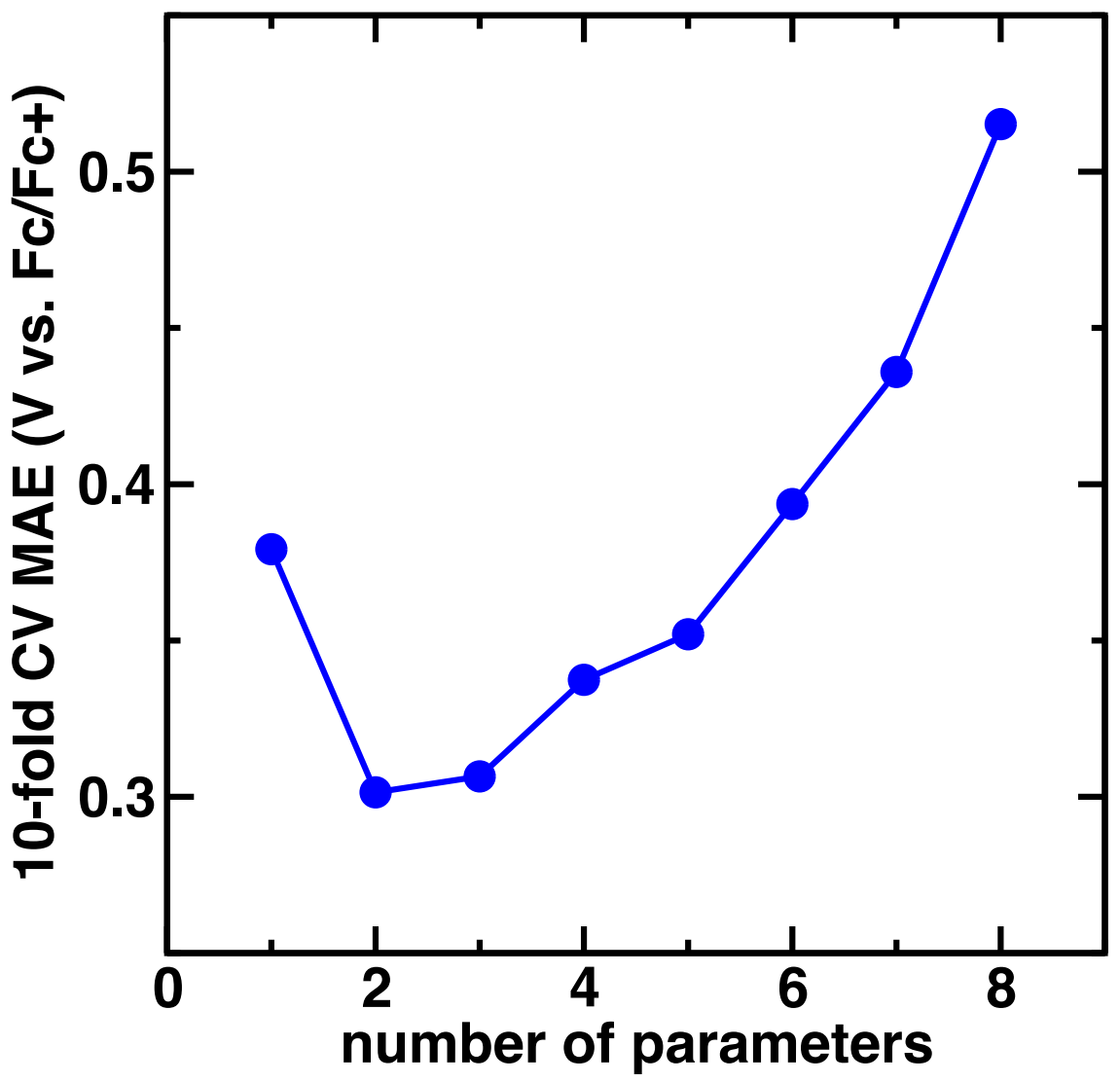

**Figure S1.** Cross-validation error for model selection on oxidation potential used for feature selection from mm-RACs on 28 heterobimetallic complexes.

**Table S4.** Train and test performance statistics for the multiple linear regression model selected by 10-fold cross-validation to predict experimental oxidation potential on 3 selected mm-RACs. The 28 complexes in the training set and 7 complexes in the test set are shown with properties reported: Pearson's *r*, mean absolute error (MAE), root mean square error (RMSE), and maximum absolute error.

| Set | Size | *r* | MAE | RMSE | max abs. error |
|---|---|---|---|---|---|
| Train | 264 | 0.89 | 0.030 | 0.043 | 0.160 |
| Test | 66 | 0.78 | 0.057 | 0.070 | 0.182 |

**Table S5.** Relative oxidation potentials (V vs. Fc/Fc$^+$ in V) for 7 complexes from literature references reports[15, 16] for complexes of the form L$^x$M2M1 where L$^X$ is the ligand type described in the main text, M1 is the upper metal, and M2 is the lower metal. The predicted value of the multiple linear regression model (i.e., test error since none of these complexes were used in model training) is also shown in V with the signed error (i.e., predicted-actual) and absolute value of the error, both in V. The largest prediction error is bolded.

| Ligand | M2 | M1 | actual | predicted | signed error | abs error |
|---|---|---|---|---|---|---|
| L$^4$ | Cr | Fe | -0.58 | -0.94 | -0.36 | 0.36 |
| L$^4$ | Cr | Co | -0.59 | -0.80 | -0.21 | 0.21 |
| L$^4$ | Cr | Cu | -0.23 | -0.39 | -0.16 | 0.16 |
| L$^4$ | Nb | Co | 0.19 | -0.44 | -0.63 | 0.63 |
| L$^4$ | Nb | Ni | -0.11 | -0.54 | -0.43 | 0.43 |
| **L$^4$** | **Nb** | **Cu** | **0.66** | **0.00** | **-0.66** | **0.66** |
| L$^4$ | Nb | Fe | -0.37 | -0.59 | -0.22 | 0.22 |

**Table S6.** Details of CSD structure curation steps

| Step | Number of results retained at each step |
|---|---|
| CSD (Conquest) search for 2 metal centers (including lanthanides, actinides, metalloids – Al/Ga/In) bonded to 3 main group elements, no polymers, R factor < 5 | 153,589 unit cells |
| Unit cells with metal-metal bonds in a bimetallic complex | 85,489 unit cells |
| Number of unique bimetallic complexes (e.g., duplicates in a unit cell) | 87,182 complexes |
| Incompatible denticities (e.g., sandwich complexes) | 62,102 complexes |
| FSR ≤ 1.5 | 38,764 complexes |
| metal CN ≤ 8 | 22,633 complexes |
| number of ligands in complex < 6 | 9,974 complexes |
| only two adjacent metals | 5,219 complexes |
| good geometry and no monodentate-only complexes | 3,684 complexes |
| FSR ≤ 1.3 | 3,493 complexes |
| **heterobimetallic (final set)** | **330 complexes** |

**Table S7.** Pauling radii used for formal shortness ratio evaluation.

| | Z | Pauling radii (Å) |
|---|---|---|
| Mg | 12 | 1.364 |
| Al | 13 | 1.248 |
| Sc | 21 | 1.439 |
| Ti | 22 | 1.324 |
| V | 23 | 1.224 |
| Cr | 24 | 1.172 |
| Mn | 25 | 1.168 |
| Fe | 26 | 1.165 |
| Co | 27 | 1.157 |
| Ni | 28 | 1.149 |
| Cu | 29 | 1.173 |
| Zn | 30 | 1.249 |
| Ga | 31 | 1.245 |
| Ge | 32 | 1.223 |
| Y | 39 | 1.616 |
| Zr | 40 | 1.454 |
| Nb | 41 | 1.342 |
| Mo | 42 | 1.291 |
| Ru | 44 | 1.241 |
| Rh | 45 | 1.247 |
| Pd | 46 | 1.278 |
| Ag | 47 | 1.339 |
| In | 49 | 1.497 |
| Sn | 50 | 1.399 |
| Sb | 51 | 1.410 |
| La | 57 | 1.69 |
| Lu | 71 | 1.557 |
| Hf | 72 | 1.442 |
| Ta | 73 | 1.343 |
| W | 74 | 1.299 |
| Os | 76 | 1.255 |
| Ir | 77 | 1.260 |
| Pt | 78 | 1.290 |
| Au | 79 | 1.336 |
| Hg | 80 | 1.440 |
| Tl | 81 | 1.549 |
| Bi | 83 | 1.520 |

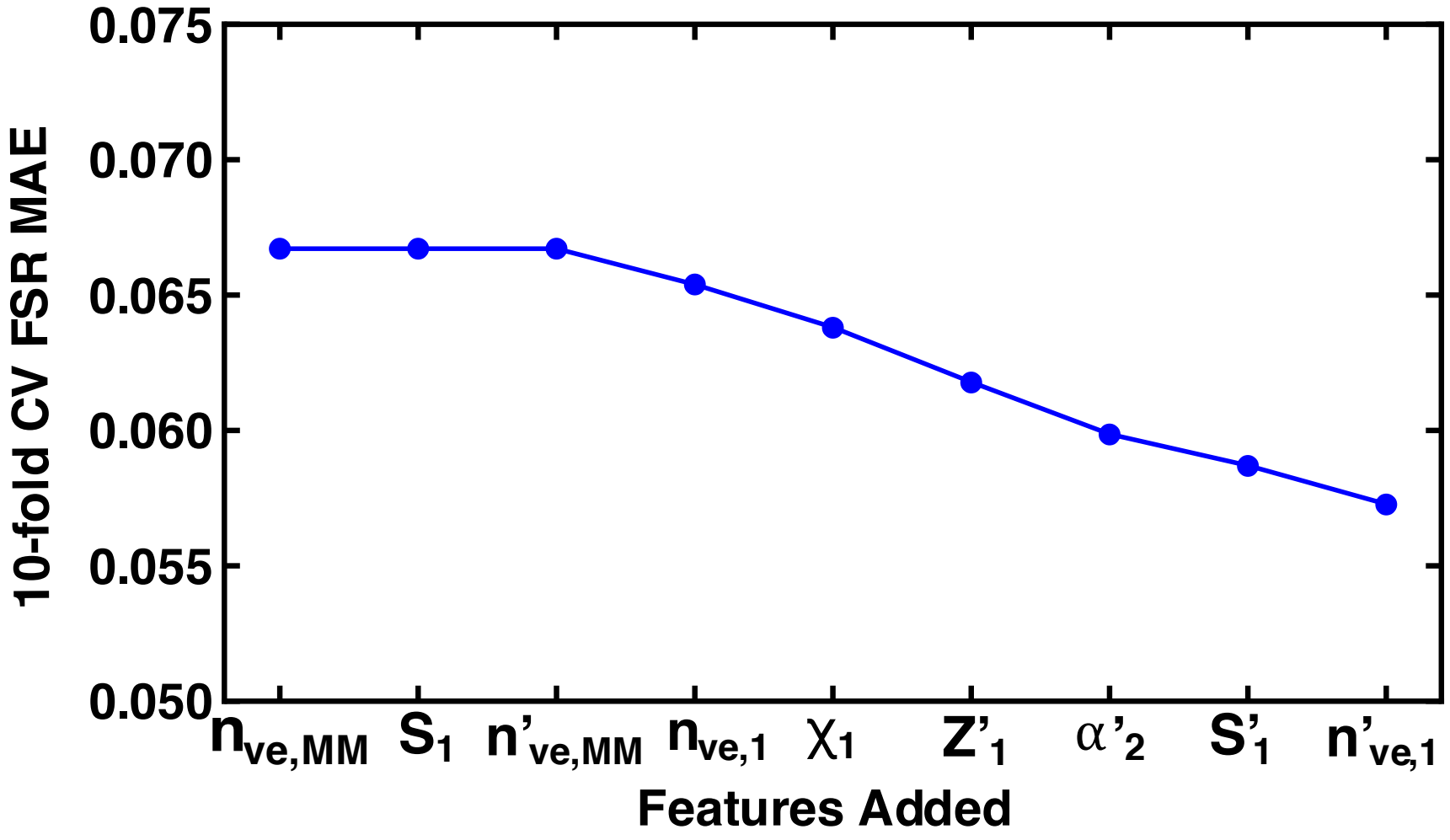

**Figure S2.** KRR model feature selection on 10-fold cross validation prediction of formal shortness ratio mean absolute error (MAE). The features are added by random-forest ranked emm-RAC features added via recursive feature addition. The above shows all 9 retained features selected in the final model for a test MAE of 0.057 on the FSR.

**Table S8.** Grid search and selected hyperparameters (α regularization and γ kernel width) for KRR model with Gaussian kernel selected on 10-fold cross-validation mean absolute error.

| | α | γ |
|---|---|---|
| log grid range | [1e-9,1] | [1e-10,10] |
| selected parameters | 5.045E-04 | 1.725E-02 |

**Table S9.** Train and test performance statistics for RF-RFA/KRR trained model to predict the FSR with 9 emm-RACs for an 80/20 split on 330 complexes: Pearson's *r*, mean absolute error (MAE), root mean square error (RMSE), and maximum absolute error.

| Set | Size | *r* | MAE | RMSE | max abs. error |
|---|---|---|---|---|---|
| Train | 28 | 0.85 | 0.26 | 0.32 | 0.73 |
| Test | 7 | 0.92 | 0.38 | 0.42 | 0.66 |

**Table S10.** Matrix of classification of FSR regimes (short, FSR < 0.9; boundary, 0.9 to 1.1; long > 1.1) for both training and test complexes. Classification disagreements are indicated in italics.

| **training data** | | | | |
|---|---|---|---|---|
| | | predicted | | |
| | | short | boundary | long |
| actual | short | 17 | *5* | 0 |
| | boundary | *4* | 198 | *14* |
| | long | 0 | *2* | 24 |
| **test data** | | | | |
| | | predicted | | |
| | | short | boundary | long |
| actual | short | 4 | *3* | 0 |
| | boundary | *3* | 42 | *3* |
| | long | 0 | *4* | 7 |

**Table S11.** Counts of 'good' metal geometry pairs for coordination environment around each metal in the retained set of 330 heterobimetallic complexes.

| **Coord env. 1** | **Coord env. 2** | **Number** |
|---|---|---|
| trigonal bipyramidal | trigonal bipyramidal | 72 |
| tetrahedral | trigonal bipyramidal | 71 |
| octahedral | square pyramidal | 59 |
| tetrahedral | tetrahedral | 38 |
| square pyramidal | square pyramidal | 20 |
| square planar | square pyramidal | 20 |
| square pyramidal | trigonal bipyramidal | 11 |
| octahedral | tetrahedral | 10 |
| octahedral | octahedral | 5 |
| seesaw | tetrahedral | 5 |
| octahedral | trigonal bipyramidal | 4 |
| square pyramidal | tetrahedral | 3 |
| square planar | square planar | 2 |
| pentagonal bipyramidal | square pyramidal | 2 |
| seesaw | trigonal bipyramidal | 2 |
| trigonal bipyramidal | trigonal prismatic | 2 |
| square planar | trigonal prismatic | 1 |
| pentagonal bipyramidal | pentagonal bipyramidal | 1 |
| tetrahedral | trigonal prismatic | 1 |
| square planar | trigonal bipyramidal | 1 |
| Total | | 330 |